# Measurement of Dielectric Suppression of Bremsstrahlung*

P.L. Anthony,[1] R. Becker-Szendy,[1] P.E. Bosted,[2] M. Cavalli-Sforza,[3,†] L.P. Keller,[1]
L.A. Kelley,[3] S.R. Klein,[3,4] G. Niemi,[1] M.L. Perl,[1] L.S. Rochester,[1] J.L. White[2]

[1] *Stanford Linear Accelerator Center, Stanford, CA 94309*

[2] *The American University, Washington, D.C. 20016*

[3] *Santa Cruz Institute for Particle Physics, University of California, Santa Cruz, CA 95064*

[4] *Lawrence Berkeley Laboratory, Berkeley, CA 94720*

## Abstract

In 1953, Ter-Mikaelian predicted that the bremsstrahlung of low energy photons in a medium is suppressed because of interactions between the produced photon and the electrons in the medium. This suppression occurs because the emission takes place over on a long distance scale, allowing for destructive interference between different instantaneous photon emission amplitudes. We present here measurements of bremsstrahlung cross sections of 200 keV to 20 MeV photons produced by 8 and 25 GeV electrons in carbon and gold targets. Our data shows that dielectric suppression occurs at the predicted level, reducing the cross section up to 75% in our data.

*Submitted to Physical Review Letters*

*Work supported by Department of Energy contracts DE-AC03-76SF00515 (SLAC), DE-AC03-76SF00098 (LBL), W-4705-ENG-48 (LLNL) and National Science Foundation grants NSF-PHY-9113428 (UCSC) and NSF-PHY-9114958 (American U.).

When a high energy electron emits a low energy photon by bremsstrahlung, the emission takes place over a long distance. In 1953, Ter-Mikaelian [1] pointed out that because of this, in a medium, bremsstrahlung of low energy photons can be suppressed. The suppression occurs because the photon production amplitude, taken over the length of the formation zone, can lose coherence because of the photon wavefunction phase shift due to the dielectric constant of the medium. This effect, known as the dielectric effect or the longitudinal density effect, suppresses the bremsstrahlung of photons with energies $k$ much less than the electron energy $E$. This suppression is important because it cuts off the bremsstrahlung photon spectrum at low energies, removing the infrared divergence in real materials. It also affects the magnitude of radiative corrections to many processes. Previous experimental work on this effect has been inconclusive [2]. We present here measurements of bremsstrahlung spectra that confirm the longitudinal density effect.

Dielectric suppression occurs because the photon emission takes place over a finite distance, known as the formation zone length. This zone may be thought of in several ways. It is the distance required for the electron and photon to separate enough (one electron Compton wavelength) to be considered separate particles. It is also the size of the virtual photon exchanged between the electron and the nucleus. Its length is given by the uncertainty principle applied to the momentum transfer between the electron and nucleus. For $k \ll E$, this momentum transfer is given by [1] [3]

$$q_\| = p_e - p'_e - k/c = \sqrt{(E/c)^2 - (mc)^2} - \sqrt{((E-k)/c)^2 - (mc)^2} - k/c \quad (1)$$

where $p_e$ and $p'_e$ are the electron momenta before and after the interaction respectively, and $m$ is the electron mass. For $E \gg m$ and $k \ll E$, this simplifies to

$$q_\| \sim \frac{m^2 c^3 k}{2E(E-k)} \sim \frac{k}{2c\gamma^2} \quad (2)$$

where $\gamma = E/m$. The formation length is then

$$l_f = \hbar/q_\| = 2\hbar c \gamma^2 / k. \quad (3)$$

If the interaction occurs in a medium, then photon interactions with the electrons in the the medium modify the relationship between the photon momentum $p$ and energy $k$ from $k = pc$ to $\sqrt{\epsilon}k = pc$ where $\epsilon$ is the dielectric constant of the medium. For energies $k$ larger than the atomic binding energies of the target electrons,

$$\epsilon(k) = 1 - \frac{k_p^2}{k^2} \tag{4}$$

where $k_p = \hbar\omega_p$ and $\omega_p = \sqrt{4\pi N Z e^2/m}$ is the plasma frequency of the medium. Here, $N$ is the number of atoms per unit volume, $Z$ is the atomic number, and $e$ is the electric charge. In particle language, the photon coherently forward Compton scatters off the electrons in the target, introducing a phase shift into the wave function. If the phase shift, accumulated over the formation zone length, is large enough, coherence is lost.

With this addition, the momentum transfer becomes

$$q_\| = p_e - p_e' - k\sqrt{\epsilon}/c \tag{5}$$
$$= \sqrt{(E/c)^2 - (mc)^2} - \sqrt{((E-k)/c)^2 - (mc)^2} - k\sqrt{\epsilon}/c \tag{6}$$
$$= \frac{k}{2c\gamma^2} + \frac{k_p^2}{2ck}. \tag{7}$$

The formation length is then

$$l_f = \frac{2\hbar c k \gamma^2}{k^2 + (\gamma k_p)^2} \tag{8}$$

Because the electron path length that can contribute coherently to a single bremsstrahlung interaction is reduced, photon emission is reduced. The emission probability is proportional to the pathlength that can contribute coherently to the emission, so the suppression S is given by the ratio of the in-material to vacuum formation lengths:

$$S = \frac{k^2}{k^2 + (\gamma k_p)^2} \tag{9}$$

For $k < \gamma k_p$, bremsstrahlung is significantly reduced. This happens for $k < rE$, where $r = k_p/m$ is a material dependent constant. For typical metals, $k_p \approx 60 - 80$ eV, so $r \sim 10^{-4}$. Table 1 gives $r$ for the targets used here.

In the absence of any suppression, the Bethe-Heitler spectrum [6] applies. It is infrared divergent with a $dN/dk \approx X_0/k$, where $X_0$ is the radiation length. For dielectric suppression, the photon spectrum is suppressed by $(k/rE)^2$, changing the Bethe Heitler spectrum to $dN/dk \approx k$.

With this suppression, the usual infrared divergence disappears, and the total cross section is finite. The total cross section depends on the target electron density and the possible presence of other suppression mechanisms; in metals it is about 10 photons per radiation length. This cutoff can also reduce the magnitude of radiative corrections involving external soft photon lines.

In addition to the longitudinal density effect, the Landau-Pomeranchuk-Migdal effect can suppress bremsstrahlung from very high energy electrons [4] [5]. Since both effects limit the formation zone length, the effects are not independent and the suppression factors cannot simply be multiplied. Migdal provided a prescription to combine the two effects [5]; his approach is used here. To minimize the contribution from LPM suppression, this analysis will concentrate on carbon targets, which exhibit relatively little LPM suppression at the energies studied here. For the targets used here, the maximum photon energies which exhibit LPM suppression in 8 and 25 GeV beams, $k_{LPM8}$ and $k_{LPM25}$, are given in Table 1.

Where LPM suppression dominates, there can be a large correction for surface interactions [7]. If an interaction occurs near a target surface, the formation zone can stick out of the target, reducing the phase shift, and hence the suppression. However, for $k < \gamma k_p$, where dielectric suppression is large, the formation zone is greatly shortened. Therefore, the dielectric effect reduces the magnitude of the 'edge effect' corrections that are required where LPM suppression is large.

The longitudinal density effect is closely related to transition radiation. Transition radiation occurs within one formation zone of the target surfaces, and has a spectrum that extends up to photon energies of $\gamma k_p$ [8]. Experimentally, the bremsstrahlung and transition radiation are indistinguishable, and can only be separated by varying target thicknesses.

We have studied the longitudinal density effect in experiment SLAC-E-146 at End Sta-tion A at the Stanford Linear Accelerator Center [9] [10]. Electrons with energies of 8 and 25 GeV entered the End Station, and interacted in several target materials. Produced photons were detected in a bismuth germanate (BGO) calorimeter 50 meters downstream, while electrons were magnetically bent downward by 39 mrad into a set of lead glass blocks that counted electrons. The one electron per pulse, 120 pulses per second electron beam was generated parasitically during SLC collider operation [11]. To minimize backgrounds, the electron path upstream of the calorimeter and the photon flight path were kept in vacuum.

TABLE I. Table I. Target $Z$, $X_0$, thickness in $X_0$, photon energy ratio for dielectric suppression, and maximum photon energies at which LPM suppression is present, for 8 and 25 GeV electron beams.

| Target | $Z$ | $X_0$ (cm) | thickness($X_0$) | $r$ | $k_{\text{LPM25}}$ (MeV) | $k_{\text{LPM8}}$ (MeV) |
|---|---|---|---|---|---|---|
| 6% $X_0$ C | 6 | 18.8 | 0.060 | $5.5 \times 10^{-5}$ | 8.5 | 0.85 |
| 2% $X_0$ C | 6 | 18.8 | 0.021 | $5.5 \times 10^{-5}$ | 8.5 | 0.85 |
| 6% $X_0$ Au | 79 | 0.34 | 0.059 | $1.1 \times 10^{-4}$ | 500 | 51.2 |

The BGO calorimeter comprises 45 crystals in a 7 by 7 array with the corners missing. Each crystal is 2 cm square and 18 $X_0$ deep. The calorimeter photomultiplier tubes (PMTs) detected about 1 photoelectron per 30 keV of energy deposition. For the data discussed here, the PMT gain was set so that 1 ADC count (250 fC) corresponded to 13 keV. The calorimeter was calibrated with cosmic ray muons, which deposited an average of 18 MeV per calorimeter crystal. The cosmic ray absolute energy scale was set by data taken with an identical cosmic ray trigger, but lower PMT gain. For this lower gain data, the absolute energy scale was determined using both a direct electron beam and with higher energy bremsstrahlung events [7]. Where bremsstrahlung data sets taken at the two gains overlapped, the agreement is good. The calorimeter temperature was monitored throughout the experiment, and the data

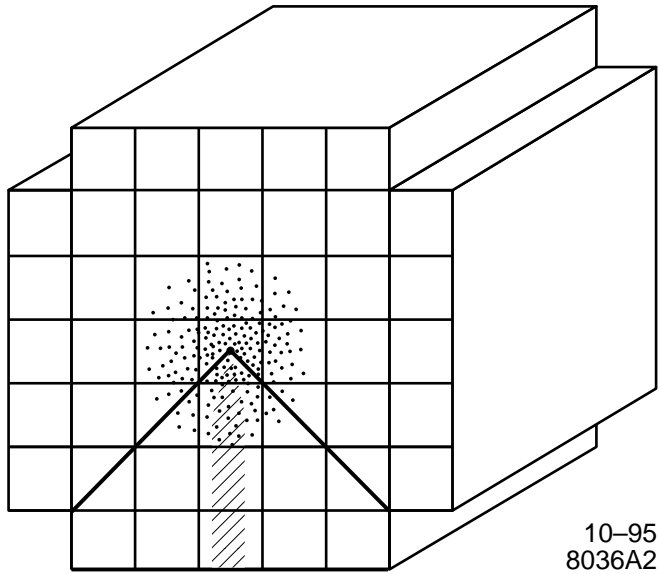

FIG. 1. Front view of the calorimeter, showing the angular cut applied. The slashes represent synchrotron radiation, while the dots represent bremsstrahlung photons.

were corrected using the measured temperature response.

Below energies of a few MeV, photon interactions in the calorimeter change character. Unlike the showers that are produced at higher energies, photons dominantly lose energy by one or more Compton scatterings. Usually, the energy loss was confined to one or two crystals in the calorimeter. To reduce background noise from synchrotron radiation and other sources, we sum only the energy in a contiguous group of calorimeter crystals. This leads to a small loss in energy for the case where a photon scatters once and then travels a long distance before its second interaction. This cluster-finding procedure, and occasional energy loss when Compton scattered photons escape from the front of the calorimeter, introduce a low-energy tail on the calorimeter response function. Because of the additional calibration step and the energy losses due to leakage, we estimate that the photon energy calibration is known to 10%.

Our analysis selected bremsstrahlung events containing a photon in the calorimeter plus a single electron in the lead glass blocks. The largest background was synchrotron radiation from the spectrometer magnet, which painted a stripe on the calorimeter, extending down-ward from the center, as shown in Figure 1. Because the magnet had a large fringe field, this background was quite small for 8 GeV electrons, with a 9 keV critical energy for electrons pointing at the bottom of the calorimeter, and an average energy deposition of 400 eV. At 25 GeV, the critical energy was 280 keV, and the average energy deposition 40 keV.

Figures 2 compares a selection of our data with the results of Monte Carlo simulations. We display histograms of the photon energy $k$ from 200 keV up to 20 MeV, plotted so that the width of the photon energy bins vary logarithmically. There are 25 bins per decade, giving each bin a width $\Delta k/k \approx 0.09$. Logarithmic bins are used so that the Bethe Heitler $1/k$ spectrum will appear as a flat line. In addition to the longitudinal density and LPM [12] effects, the simulation [7] includes the effects of multiple photon emission, where one electron undergoes two independent bremsstrahlung interactions in the course of passing through the target. This effect changes the slope of the Monte Carlo curves. The simulation also includes transition radiation [8] and allows for the possibility that produced photons might interact in the target via either pair-production or Compton scattering. Finally, the code includes a simple simulation of the calorimeter resolution.

Figure 2a shows the data for an 8 GeV beam passing through 6% $X_0$ of carbon. Photons from 200 keV to 20 MeV are included, corresponding to $0.22r < k/E < 22r$. Three predictions are shown: Bethe Heitler, a curve with the LPM effect only, and a curve that includes both LPM suppression and the longitudinal density effect. Here, the LPM effect is small. Only the LPM plus longitudinal density effect curve fits the data; the other curves are strongly excluded.

Figure 2b shows the data for an 8 GeV beam passing through 2% $X_0$ of carbon, with the same Monte Carlo curves. The data and Monte Carlo curves are flatter than in Figure 2a because of the reduced multi-photon pileup. Because of the thinner target, transition radiation is more visible; it causes the upturn visible in the Monte Carlo and data below 300 keV. Again, only the LPM plus longitudinal density curve fits the data. The good agreement between the data and theory leaves little room for additional emission at the target surfaces.

Figure 2c shows data for an 8 GeV beam passing through 6% $X_0$ of gold. The 200 keV

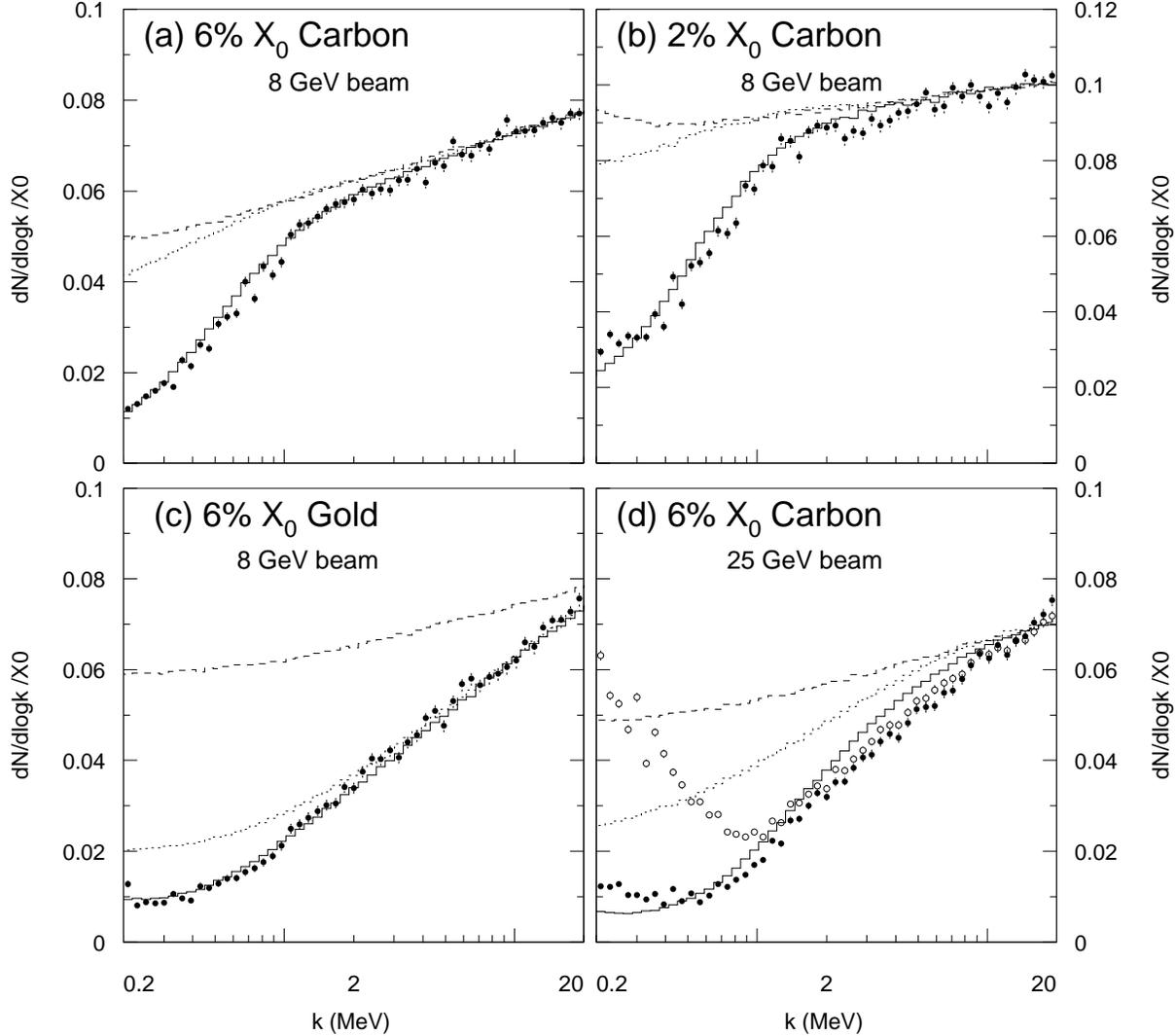

FIG. 2. Measurements (with statistical errors only) of the photon spectrum $dN/d(\log k)$ from 200 keV to 20 MeV compared with Monte Carlo calculated theoretical curves The cross sections are given in terms of $dN/d(\log k)/X_0$ where $N$ is the number of photons per energy bin per incident electron. The photon energy scale is logarithmic with 25 bins per decade, so each bin has a width $\Delta k \sim 0.0964k$. The dashed histogram is the Bethe-Heitler Monte Carlo, the short dashes show the LPM only prediction and the solid histogram is the LPM plus longitudinal density effect calculation. The figures are: (a) 8 GeV electrons incident on 6% $X_0$ carbon; (b) 8 GeV electrons incident on 2% $X_0$ carbon. (c) 8 GeV electrons incident on 6% $X_0$ gold; (d) 25 GeV electrons incident on 6% $X_0$ carbon. For the latter, the open and filled circles represent the data before and after the angular cut is applied, respectively.

to 20 MeV photon energy range covers $0.11r < k/E < 11r$. Here, LPM suppression is larger than the longitudinal density effect. The curve with both effects is strongly preferred. A prediction based on simply multiplying the two suppressions together, rather than combining them as was done here, would be far below the data. Because gold is denser than carbon, the transition radiation is larger than in the previous figures, accounting for about 60% of the LPM plus dielectric effect cross section at 200 keV. The predicted total emission is close to a minimum around 200 keV; at lower energies the transition radiation rises sharply.

Figure 2d shows the 25 GeV data from the 6% $X_0$ carbon target. The photon energy range is unchanged; it corresponds to $0.07r < k/E < 7r$. At the higher beam energy, synchrotron radiation is a large background below about 1 MeV. Because of this background, we show two sets of experimental points, one raw (open circles), and the other with a cut which removes most of the synchrotron radiation (solid circles). The cut removes photons centered in the bottom quarter of the calorimeter, below the diagonal lines in Figure 1. This cut removes synchrotron radiation, indicated by the hatching in Figure 1, while leaving 75% of the bremsstrahlung signal. Photons on the borderlines were given an appropriate weighting factor. The cut efficiency is independent of the photon emission angle. The independence is important because the suppression is expected to disappear for photons emitted at angles larger than $1/\gamma$. Unfortunately, the efficiency of this cut depends on how well the beam was centered on the calorimeter. The average deviation from the calorimeter center, was less than 0.5 cm, corresponding to a 15% systematic error. Data from no-target runs shows that the cut removes about 80% of the background.

With the cut, the data and LPM plus longitudinal density effect Monte Carlo are in good agreement down to about 500 keV; below this energy an excess remains, consistent with the expected cut efficiency. Below about 400 keV, almost all the target induced signal is expected to be from transition radiation. This plot shows that the dielectric suppression scales with energy as expected, and further demonstrates that Migdal's method [5] for combining the longitudinal density effect and LPM effect works.

The Monte Carlo curves are normalized to match the data, using normalization constants found at photon energies of 5 to 500 MeV [7]. In most cases, the data is slightly above the Monte Carlo predictions; the average shift was about 6%. Except for 25 GeV electrons incident on the carbon targets, the normalizations found using the data presented here match those found at higher photon energies. For the carbon targets in 25 GeV beams, below photon energies of about 10 MeV, the intensity appears lower than the Monte Carlo. Because this may be due to the target material structure, we use the normalization found at higher photon energies here.

The points show statistical errors only. The major systematic errors which can vary with energy are due to: photon cluster finding (7%), calorimeter nonlinearity (3%), overall energy calibration (3%), remaining backgrounds (4%), target density uncertainty (2%) and Monte Carlo inadequacies, mostly in handling the multiphoton pileup (1%). Added in quadrature, these give a total systematic error of 9%. Data which includes the angular cut has an additional 15% systematic error.

In conclusion, we observe that the emission of bremsstrahlung of photons with energies 200 keV to 20 MeV from 8 and 25 GeV electrons is suppressed as predicted by the longitudinal density effect. The effect shows the expected energy dependence, and the magnitude is within 10% of that expected. Where both the longitudinal density effect and LPM suppression are present, they combine as predicted by Migdal.

We would like to thank the SLAC Experimental Facilities group for their assistance in setting up the experiment and measuring the magnetic field, the SLAC Accelerator Operators group for their efficient beam delivery.

# REFERENCES

† Present address: Institut de Fisica d'Altes Energies, Universitat Autonima de Barcelona, 08193 Bellaterra (Barcelona), Spain.

[1] M. L. Ter-Mikaelian, Dokl. Akad. Nauk. SSR **94**, 1033 (1954). For a discussion in English, see M. L. Ter-Mikaelian, *High Energy Electromagnetic Processes in Condensed Media*, John Wiley & Sons, 1972.

[2] F. R. Arutyunyan, A. A. Nazaryan and A.A. Frangyan, Sov. Phys. JETP **35**, 1067 (1972).

[3] V. M. Galitsky and I. I. Gurevich, Il Nuovo Cimento **32**, 396 (1964).

[4] L. D. Landau and I. J. Pomeranchuk, Dokl. Akad. Nauk. SSSR **92**, 535 (1953); **92**, 735 (1953). These two papers are available in English in L. Landau, *The Collected Papers of L. D. Landau*, Pergamon Press, 1965.

[5] A. B. Migdal, Phys. Rev. **103**, 1811 (1956).

[6] H. A. Bethe and W. Heitler, Proc. Royal Soc. A **146**, 83 (1934).

[7] P. Anthony *et al.*, Phys. Rev. Lett. **75**, 1949 (1995).

[8] J. D. Jackson, *Classical Electrodynamics, 2nd edition*, John Wiley & Sons, 1975, pg. 685-693.

[9] S. R. Klein *et al.*, in *Proc. XVI Int. Symp. Lepton and Photon Interactions at High Energies* (Ithaca, 1993), Eds. P. Drell and D. Rubin, p. 172.

[10] R. Becker-Szendy *et al.*, in *Proc. 21st SLAC Summer Institute on Particle Physics* (Palo Alto, 1994), pg. 519.

[11] M. Cavalli-Sforza *et al.*, IEEE Trans. Nucl. Sci. **41**, 1374 (1994).

[12] T. Stanev *et al.*, Phys. Rev. D **25**, 1291 (1982).